\documentstyle[11pt,aaspp4]{article}
\def\etal{{et al.}\thinspace}
\def\beb{\begin{thebibliography}{999}}
\def\eeb{\end{thebibliography}}
\def\bi{\bibitem[]{}}
\def\be{\begin{equation}}
\def\ee{\end{equation}}
\def\bea{\begin{eqnarray}}
\def\eea{\end{eqnarray}}

\newcommand{\gsim}{\raisebox{-0.3ex}{\mbox{$\stackrel{>}{_\sim} \,$}}}
\newcommand{\lsim}{\raisebox{-0.3ex}{\mbox{$\stackrel{<}{_\sim} \,$}}}
\begin{document}
\date{}

\title{Sunyaev-Zel'dovich distortion from early galactic winds}

%\maketitle
\author{Subhabrata Majumdar$^{1,2}$ , Biman B. Nath${^3}$ 
\& Masashi Chiba${^4}$}

\affil{$^1$Joint Astronomy Programme, Physics Department,\\
Indian Institute of Science, Bangalore 560012, India}
\affil{$^2$Indian Institute of Astrophysics, Koramangala, Bangalore 560034,
 India}
\affil{$^3$Raman Research Institute, Bangalore 560080, India}
\affil{$^4$National Astronomical Observatory, Mitaka, Tokyo, Japan }
\bigskip\bigskip

\begin{abstract}
We consider the distortion in the cosmic microwave background (CMB) due to
galactic winds at high redshift.
Winds outflowing from galaxies have been hypothesized to be
possible sources of metals in the intergalactic medium, which is known
to have been enriched to $10^{-2.5}$ Z$_{\odot}$ at $z \sim 3$. We
model these winds as functions of mass of the parent galaxy and redshift, 
assuming they activate at a common initial redshift, $z_{in}$, and calculate the mean
$y$ distortion and the angular power spectrum of the distortion in
the CMB. We find that the thermal Sunyaev-Zel'dovich (SZ) effect due to 
the winds to be consistent with previous estimates. The distortion due to 
the kinetic SZ (kSZ) effect is, however, found to be
more important than the thermal SZ (tSZ) effect. 
%Unlike in the case of SZ distortion from clusters of galaxies, we find that 
%the upper limits set by recent  measurements
%of CMB temperature variance at arc minute scales cannot be used to constrain 
%models of galactic winds. It is, however,
%clear 
We find that the distortion due to galactic 
winds is an important contribution to the power spectrum of distortion
at very small angular scales $(l \sim 10^4 )$.
We also find that the power spectrum due to clustering dominates the
Poisson power spectrum for $l \le 4\hbox{--}5 \times 10^5$. We show 
explicitly how the combined power spectrum from wind dominates over that
of clusters at $217$ GHz, relevant for PLANCK.
We also show how these constraints change when the efficiency of the winds
is varied.

\end{abstract}

\keywords{Cosmology: cosmic microwave background, galaxies : formation,
galaxies : intergalactic medium}

\section{Introduction}

Recent studies of the epoch of reionization of the universe
offer the following scenario for the first baryonic objects.
Although the first objects form at redshifts $z \sim 40$ with
masses of order $10^5$ M$_{\odot}$, UV radiation from stars formed in these
objects obliterate the molecular hydrogen in other objects in
the vicinity, thereby suppressing further formation of objects
with masses $M \le 10^8$ M$_{\odot}$ (Haiman, Rees \& Loeb 1997;
Ciardi \etal 1999). These objects with masses comparable to
local dwarf galaxies appear later, at redshifts
$z \lsim 15$.

These galaxies are thought to have affected the intergalactic
medium (IGM) in at least two ways. Along with the radiation from
early black holes, they provided the UV radiation needed to reionize
the IGM and keep it ionized thereafter. It is also possible that outflows
from them provided the metals needed to explain the enrichment
of the IGM (Miralda-Escud\'e \& Rees 1997; Nath \& Trentham 1997;
Aguirre \etal 2000). 
Observations show that the metallicity of the IGM
at $z \sim 3$ was $\sim 10^{-2.5}$ Z$_{\odot}$ (Songila \& Cowie
1996; Songaila 1997). Although other mechanisms, like galactic
mergers, could also have been important in enriching the IGM
(Gnedin \& Ostriker 1997), here we consider only 
galactic winds.

Dwarf starbursting galaxies are expected to produce energetic
outflows driven by supernovae and winds from OB stars in them
(Larson 1974; Dekel \& Silk 1986). This has been observed both
locally (e.g., Heckman 1997) and at high redshift (Franx
\etal 1997; Pettini \etal 1998). It is not, however, entirely
clear what the efficiency of these outflows is in driving out
the gas in the galaxy. Mac Low \& Ferrara (1999) have recently
shown that efficient blowaway occurs only in halos with masses
$M \le 10^7$ M$_{\odot}$, and that a blowout occurs perpendicular
to the disk for $10^8 \le M \le 10^9$ M$_{\odot}$, although 
their simulations assume a disk geometry. But it is certain that
galactic outflows do occur and have important effects on the
ambient medium. 

Recent studies on galactic outflows at high redshift contend
that these winds could have suppressed galaxy formation in
the vicinity, for masses $\le 10^9$ M$_{\odot}$ by 
mass stripping and gas heating (Scannapieco \&
Broadhurst 2000). They assume spherical shells of winds from
dwarf spheroidals at high redshift. They also find that these
outflows could have distorted the microwave background to
an extent that is marginally below the current detection limits,
with a mean Compton parameter $\sim 10^{-5}$ depending on
the model parameters. Although they did not explicitly calculate
the power spectrum of anisotropy in the microwave background, 
it is clear that distortion from these
outflows can be important and deserves detailed calculations
and comparison with distortion from other phenomena. 

This is what calculate 
in this paper. In addition, we take into account the effect of kinetic
SZ effect and the effect of clustering of parent galaxies. The importance
of the kinetic SZ effect of outflows---in the case of quasar outflows---has 
already been shown by
Natarajan, Sigurdsson \& Silk (1998) and Natarajan \& Sigurdsson (1999).
Here, We consider in detail the effect of the outflows from
high redshift dwarf spheroidals on the cosmic microwave background.
We calculate the properties of the outflows and then the angular
power spectrum of the distortions. We consider the effect on the
CMB from outflows from a single initial redshift, $z_{in}$. Finally, we
vary this initial redshift and determine the dependence of the
distortion on it.

The paper is structured as follows. The next section discusses the
characteristics of the outflows, and then in \S 3, we set up an
ensemble of these outflows. In \S 4, we introduce the formalism for
calculation of the Sunyaev-Zel'dovich effect from these outflows, and we
present the results in \S 5. We end with a discussion of these results
in view of several observational constraints.

We assume a flat universe with a cosmological constant, with $\Omega_0=0.35$,
$\Omega_{\Lambda}=0.65$ and $h=0.65$ as our fiducial model. 

\bigskip

\section{Modelling the cosmological outflows}

We model the outflows following the approach of Tegmark, Silk
\& Evrard (1993; hereafter TSE). They considered the evolution
of the blastwaves with an initial energy input (which stops 
after $t_{burn} \sim 5 \times 10^7$ yr), which have thin
shells sweeping up most of the ambient IGM gas.  The shells
are assumed to lose a small fraction $f_m=0.1$ of their mass to
the interiors.

We assume, as in TSE, that the total mechanical luminosity
of the outflow is $L_{sn} \sim 1.2 (M_b/M_\odot) L_\odot $, where
$M_b$ is the baryonic mass of the parent galaxy. Since the galactic winds
last for $t_{burn} \sim 5 \times 10^7$ yr, 
the total energy in the galactic wind is then
$0.02 \times 0.007 \times M_b c^2$, where $M_b$ is the baryonic mass in the
galaxy. The factor of $0.02$ follows from the observations of Heckman, Armus,
\& Miley (1990) who show
that approximately 2\% of the total luminosity of high redshift starburst galaxies goes into
galactic winds. If one assumes the galactic wind to have a constant mechanical
luminosity $L_{sn}$ for the duration of $t_{burn}$, then TSE show that
$L_{sn} \sim 1.2 (M_b/M_\odot) L_\odot$. 

To determine the baryonic fraction, we can either use the baryonic
fraction in clusters, or the baryonic fraction of matter in the
diffuse IGM, using $\Omega_b = 0.05$. The last fraction, for a low density
universe, with $\Omega_0=0.35$ as in our model, corresponds to $\sim 0.15$.
Also, for clusters, White et al. (1993) has estimated the baryon 
content to be
$M_b/M \ge 0.009+0.05h^{-3/2}$. Adopting the Hubble parameter $h=0.65$,
gives the baryon gas mass fraction to be be $\sim 0.1$.
For these reasons, in this
calculation we assume that $M_b=0.1 \> M$. We also, however, show the 
final results for the assumption $M_b=0.05 \> M$, to show the effect of
such changes on the final constraints. Since the mechanical luminosity
in this model is proportional to $M_b$, this will show the effect of
changing the efficiency of winds on our results. 

We then numerically
solve for the radius and velocity of the shell, and the
energy and particle density inside the shell. For the radius and velocity
of the shells, we use the equation 1 of TSE, that is,
\begin{equation}
\ddot R= {8 \pi p G \over \Omega_{IGM} H^2 R} - {3 \over R} (\dot R -HR)^2
-  ( \Omega_d + 0.5 \Omega_{IGM}) (0.5 H^2 R) \>.
\end{equation}
Here, $\Omega_{IGM}$ is the ratio of the density of the intergalactic
gas to the critical density, and $\Omega_d$ expresses the total density.
The evolution
of the total energy inside the shell is given by TSE as,
\begin{equation}
{d E_t \over dt}= L_{sn}-4 \pi p R^2 \dot R - L_{brem} -L_{comp}.
\end{equation}
Here, the first term refers to the energy input from the galaxy,
the second term describes the adiabatic loss, the third term describes
the energy loss due to Bremsstrahlung and the last term is
the energy loss rate due to inverse Compton scattering off the microwave
background photons, which becomes important at high redshifts. 
$L_{brem}$ is found to be always much smaller than
$L_{comp}$, as was noted by TSE. We have here neglected the contribution
from the somewhat
uncertain source of heating to collision of
the shell with the IGM from ($L_{diss}$ in TSE) to
obtain a conservative estimate of the CMB distortion. 

If the shells lose a fraction $f_m$ to the interiors, then the particle
density inside the shell of radius $R$ evolves as (assuming a uniform
density),
\begin{equation}
{dn \over dt}=f_m {d \rho _{IGM} \over dt} + {3 \over R} {dR \over dz}
\, f_m \rho _{IGM} \,,
\end{equation}
where $\rho_{IGM}$ is the mass density of the ambient IGM at the given
redshift. 

For initial conditions, we assume that the initial radius of the shell
is the radius of the dwarf spheroidal, for which we use the scaling
of the optical size of elliptical galaxies being $=1.2 \times 10^2 (M/ 10^{12} 
M_{\odot})^{0.55}$ kpc (Saito 1979; Matteucci \& Tornambre 1987). 
This is a useful relation, although its validity has not been updated
with more recent data. Our wind solutions, however, do not depend strongly
on the assumption of this initial point, and therefore its exact validity
is not relevant here.
The initial velocity is expected
to be of the order of the thermal velocity of gas heated to $10^6$ K.
We assume here that the blastwaves expand into an already ionized
medium, and therefore, neglect the energy loss due to ionization of the
IGM (as considered by TSE). The early supernovae driven winds are
expected to expand into an IGM, which has already been ionized by
the same massive stars which gave rise to the supernovae later 
(Ferrara, Pettini \& Shchekinov 2000). We assume that $\Omega_{IGM}=\Omega_b
=0.05$
as constrained by big bang nucleosynthesis (for $h=0.65$), 
since at redshift $z \sim 4\hbox{--}5$ most of the
baryons are in the IGM (Rauch \etal 1998; Weinberg \etal 1997). 

With the above equations and initial conditions we calculate the wind
parameters, for a given initial redshift $z_{in}$, as functions of the
mass of the parent galaxy and redshift, using a fourth order Runge-Kutta
technique. We plot in Figure 1 the evolution of the radius (in physical 
coordinates)
and the electron temperature for winds out of galaxies of masses,
$M=10^9$ and $10^{10}$ M$_{\odot}$ for $1+z_{in}=15,20$. We have checked
that our results for radius and $T_e$ match with that of Scannapieco \&
Broadhurst (2000) for similar parameters (their Figure 3).

We note here that the uncertainty in the efficiency of mass loss in
galactic outflows, as questioned by Mac Low and Ferrara (1999), is
not relevant here, as we assume that the entire mass in and inside the
shell is provided by the ambient IGM. We only use the mechanical
energy output of the parent galaxy in our calculations.

\bigskip

\section{Ensemble of galactic outflows}

Although a series of outflows
would probably occur in the universe, for simplification and
tractability of the calculation, we consider the case of outflows
originating at a single epoch, at $z_{in}$. We consider a large range
of values for $3 \le z_{in} < 15$. This assumption of a single epoch
can be justified
to some extent because of the fact that after the first objects
of mass of order $\sim 10^8$ M$_{\odot}$ start shining and outflowing,
the resulting UV background radiation and the winds both inhibit
the further formation of low mass objects: photons do this by heating up
the IGM gas (Thoul \& Weinberg 1996), 
and the outflows, by gas stripping (Ferrara, Pettini
\& Shchekinov 2000). Therefore, if there were widespread outflows
associated with the epoch of first luminous objects with masses
of order $10^8$ M$_{\odot}$, it could not have lasted a long time.

We use the results of the previous section to set up an ensemble of 
galactic outflows from galaxies with masses $5 \times 10^7 \le M \le 10^{11}$
M$_{\odot}$, using the abundance of collapsed objects as predicted
by a modified version of the  Press-Schechter (PS) mass function 
(Press \& Schechter 1974) given by Sheth and Tormen (1999) which
matches N-body simulations well at galactic scales.
We assume that the galaxy number density is linearly biased and
traces the abundance of collapsed dark matter halos.
The lower limit of mass of parent galaxies is motivated by the fact
that the first baryonic objects to shine after the initial pause 
(due to destruction of molecular hydrogen) have virial temperature
of order $10^4$ K, with corresponding masses of order
$5 \times 10^7$ M$_{\odot}$ (Haiman, Rees \& Loeb 1997;
Scannapieco \& Broadhurst 2000). The upper limit of
masses is motivated by the fact that a blowout seems to be inhibited
for masses above $10^{12} (1+z)^{-3/2} \, M_{\odot}$, as reported
by Ferrara, Pettini \& Shchekinov (2000), although we note that these
limits apply strictly to disk galaxies, whereas we consider here
dwarf spheroidals as parent galaxies.

The PS mass function essentially calculates the mass function of 
objects in terms of their collapse redshift $z_{c}$. In principle,
the epoch of galactic winds will occur later, and so, in principle,
$z_{in} \le z_{col}$. The fractional difference between these two epochs
is, however, extremely small. There are two effects that can make $z_{in}$
differ from $z_c$ in principle: the extra time for cooling and then the time
for galactic winds. Firstly, the change in the PS formalism
due the effect of cooling, from $(1+z_c)$ to $(1+z_c)(1+M/M_{cool})^{2/3}$
is negligible here, as $M_{cool} \sim 3.6 \times 10^{11} 
 \> M_{\odot}$, (Peacock \& Heavens 1990)
larger than the objects considered here.
Moreover, the timescale for the galactic wind for dwarf galaxies
is much smaller than a Hubble time at $z\sim 15$ (Nath \& Trentham 1997).
We, therefore, assume that $z_{in}=z_c$.

We use the transfer function of Bardeen \etal (1986)
with the shape parameter given by Sugiyama (1995), and the Harrison -
Zel'dovich primordial spectrum to calculate the matter power spectrum
$P_m$(k). The resulting mass  variance ($\sigma_8$) on scales of
$8h^{-1}$ Mpc is normalised to 4-year $COBE-FIRAS$ data using the fit given
by Bunn \& White (1997). This normalisation, unlike that from cluster
abundance, is free of uncertainties of physics of cluster formation. 

%We use an initial power law spectrum with an effective spectral index
%$n = -2$ to model structures of galactic scales. The present mass
%variance for the power law spectrum is given by
%$\sigma(M)= \sigma_{8} {\left(M/M_8\right)}^{-\alpha}$, where 
%$\alpha = (n+3)/6$ , $M_8 = 6\times10^{14} \Omega_0 h^{-1}M_{\odot}$
%is the mass within $R_8 = 8h^{-1}$Mpc and $\sigma_8$ is the
%normalisation (Lacey \& Cole 1993). Following Viana \& Liddle (1996)
%we have calculated the cluster normalised $\sigma_8$ for a particular 
%set of cosmological parameters. 

Once the objects are distributed,
we then calculate the peculiar velocities with respect to the CMB frame
of reference. The velocities are usually assumed to follow a Gaussian
(Moscardini \etal 1996, Bahcall \etal 1994, Bahcall \&
Oh 1996), or Maxwellian distribution (Molnar \& Birkinshaw 2000),
which is completely defined by its rms value $\sigma_v$. For analytical
simplicity we assume a mean peculiar velocity for all 
galaxies for a given redshift
and cosmology, which is equal to the rms value $\sigma_v$ of the distribution.
Following Molnar and Birkinshaw (2000), we can write the cluster peculiar
velocity $v_r$ expected from a Gaussian initial density field as as Maxwellian
distribution with a pdf given by ${v_r}^2 exp[-{v_r}^2/2\sigma_v(z)^2]$. 
The rms peculiar velocity is then given by
$<v^2>_R(z)=H^2(z)a^2(z)f^2(\Omega_0,\Lambda)\sigma_{-1}(R)$, where $a(z)$ is
the scale factor and $\sigma_{-1}(R) = 1/{2\pi^2}{\int_0^\infty} P(k)W(kR)dk$.
Here $W_R$ is the top-hat window function for smoothing. The velocity factor
is $f(z)\equiv d$ln$\delta/d$ln$a$ (see Lahav et al. 1990, Peebles 1980 for 
detailed expression of $f(z)$ ). From averaging over the above Maxwellian  
pdf, one can get the Maxwellian width given by
\be
< v_r^2> = {{\int_0^\infty v^4 exp[-{v_r}^2/2\sigma_v(z)^2]dv}\over
{\int_0^\infty v^2 exp[-{v_r}^2/2\sigma_v(z)^2]dv}} = 3\sigma_v^2.
\ee
Next, one expresses $\sigma_v$ as Normalisation $\times
[H(z)a(z)f(z)]/[H(0)a(0)f(0)]$. The normalisation at $z=0$ is taken to be
$v_{r0}= 400 km s^{-1}$ for flat models with a cosmological constant (Gramann,
1998). But we also show our results with a smaller $v_r(z=0) = 300$ km s$^{-1}$.
For the $\Lambda$CDM model we can write $\sigma_v$ explicitly (Molanar
and Birkinshaw, 2000) as
\be
\sigma_v(\Omega_0,\Lambda,z) = 400 km s^{-1}{{1+z}\over{{[\Omega_0(1+z)^2 + 1
-\Omega_0]}^{1/2}}} {{D_{\Lambda}(\Omega_0,0)}\over{D_{\Lambda}(\Omega_0,z) }}
\left[{{5-3(1+z)D_{\Lambda}(\Omega_0,z)}\over{5-3D_{\Lambda}(\Omega_0,0)}}
\right]
\ee
In the above formula, $ D_{\Lambda}(\Omega_0,z)$ is the linear growth factor
(see Caroll,Press \& Turner 1992 for approximate expression for a $\Lambda$CDM 
universe). Given a large number of realisations of a given cosmology, we expect
the final result to be close to that obtained from using the mean velocity.

\bigskip
\section{Sunyaev-Zel'dovich distortion}

The hot gas in the interior of the shells can distort the microwave
background, introducing temperature anisotropies, through the 
Sunyaev-Zel'dovich effect (Sunyaev \& Zel'dovich  1972).
Scattering of the photons off hot electrons causes the radiation to gain
energy. Also conservation of photons in the scattering results in a
systematic shift in photon number from the R-J to the Wein side of the
spectrum. The spatial dependence of the effect is given by the Comptonization
parameter $y= 
2 \int_0 ^R {k_B T_e \over m_e c^2} \sigma _T n_e dl 
$
where $n_e, \, T_e$ refer to the electron density and temperature
inside the shell of radius $R$ (physical size), $\sigma_T$ is the
Thomson cross-section, $k_B$ is the Boltzmann constant, and $m_e$ is the
electron mass. The thermal SZ effect is characterised by its specific
spectral signature. The change of spectral intensity $I$, along the line of
sight is given by $\Delta I_{th} = y i_0  g(x)$, with
$i_0 =  2{(k_bT_0)}^{3}/{(hc)}^{2} $.
The spectral form of the thermal SZ effect is given by
\be
g(x) = {{x^4 e^{x}}\over{{(e^{x} - 1)}^2}}
\left[ x coth(x/2) - 4 \right] ,
\ee
where $x=h\nu/K_bT_0$. In the non-relativistic limit the spectral function is
zero at the crossover frequency $x_0 = 3.83$ (i.e. $\nu = 217$ GHz with 
$T_0 = 2.726$). The frequency  dependence of the corresponding temperature 
change due to tSZ is given by
\be
{{\Delta T}\over{T_0}} = \left[ x coth(x/2) - 4 \right] y.
\ee
If $x<<1$ (R-J limit), then $\Delta T/T_0 = -2y$, and for $x>>1, 
\Delta T/T_0 = x^2y$.
This specific spectral dependence of the thermal SZ effect can be used to
separate it out from the primary anisotropy. 
The thermal SZ effect is caused by the random thermal motion of the electrons
whose distribution is isotropic (in the wind frame).

An additional source of anisotropy is due to the 
Doppler
effect of the CMB photons if the shell has a peculiar velocity. For
a radial peculiar velocity $v_r$ of the shell (w.r.t the Hubble flow)
, the amplitude of the
kinetic SZ effect is given by,
\be
{\delta T \over T}={v_r \over c} \times  \int_{-R} ^R \sigma _T
n_e dl \,.
\ee
 The intensity change due to the peculiar motion is 
$ \Delta I_{kin} = - {{v_r}\over{c}}\tau  i_0 h(x)$, where $\tau$ is the
cluster optical depth to Compton scattering, and 
\be
h(x) = {{x^4 e^{x}}\over{{(e^{x} - 1)}^2}} .
\ee
However, in contrast to with the temperature change due to tSZ, the kinematic
temperature change due to kSZ is independent of frequency  (see Rephaeli,
1995). Thus one can utilise the different frequency dependence of temperature
anisotropy of tSZ and kSZ to separate them out in any observation involving
multiple frequency.

%The kinetic SZ effect has the same spectral signature as that of the
%primary CMB anisotropies, and hence is difficult to separate from
%primary anisotropies. 

The ratio of thermal SZ effect to the kinetic SZ effect is then given by
${k_B T_e \over m_e c^2} /{2 v_r \over c}$.
It is generally found that larger structures (like clusters of galaxies), 
which form later,
will have a larger thermal SZ effect than
kinetic SZ effect, whereas structures going back to high redshifts (like
quasar ionised bubbles, black hole seeded proto-galaxies: see Aghanim \etal
1996, 2000; Natarajan \& Sigurdsson 1998) will have the opposite behaviour. CMB distortions due to galactic
winds fall in the second category.

For simplicity we assume that the density and temperature 
inside the shells are uniform. At high redshift, this is a realistic
approximation as the sound crossing time is less than the Hubble time.
At lower redshifts, this approximation admittedly breaks down
to some extent. 

The fluctuations of the CMB temperature produced by either the thermal
or the kinetic SZ effect  can be quantified by their spherical harmonic
coefficients $a_{lm}$ , which can be defined as
$\Delta T({\bf n}) = T_0^{-1} \sum_{lm} a_{lm} Y_{lm}({\bf n})$. The
angular power spectrum of SZ effect is then given by
$C_l=<{|a_{lm}|}^2>$, the brackets denoting an ensemble average. We
first consider the objects to be Poisson random distributed,
and consider the correlation betweeen them later.
The power spectrum for the
cosmological distribution of objects, can then be written as 
(Cole \& Kaiser 1988, Peebles 1980)
\be
C_l^{Poisson} = \int_0^{z_{max}} dz {{dV(z)}\over{dZ}} \int_{M_{min}}^{M_{max}}
     dM {{dn(M,z_{in})}\over{dM}} {|y_l(M,z)|}^2
\ee	 
where $V(z)$ is the comoving volume and $dn/dM$ 
is the number density of objects.
Following Kitayama and Komatsu, we define the power spectrum of y-parameter
which is independent of frequency as $C_l = C_l(x)/g^2(x)$ with g(x) given by
eqn (4) for tSZ. For kSZ $C_l=C_l(x)$.
(Note that $z_{max} \le z_{in}$ in our case.) In our calculation, we use
a modification of the Press-Schecter distribution, which provides a
good fit to the unconditional halo mass function for different cosmologies
in the mass range of our interest (the details can be found in Sheth \& Tormen
(1999)). 
Since these fluctuations occur at
very small angular scales, we can use the small angle approximation of Legendre
transformation and write $y_l$ as the angular Fourier transform of $y(\theta)$.
as  $y_l = 2\pi \int_{r_{gal}}^{r_{wind}} y(\theta) J_0 [(l+1/2])\theta]\theta
d\theta $. We have used the approximation $P_l\approx J_0
[(l+1/2])\theta]$
, where $P_l$ is a Legendre coefficient and $J_0$ is a Bessel function of first
kind and zero order (Peebles 1980, Molnar \& Birkinshaw 2000). 

At high $z$ we 
expect a significant contribution to the anisotropy  from correlation among the
structures. Following Komatsu and Kitayama (1999), 
we estimate the angular 
power spectrum as
\be
C_l^{Clustering} = \int_0^{z_{max}} dz {{dV(z)}\over{dZ}} P_m
{\left[\int_{M_{min}}^{M_{max}} dM {{dn(M,z_{in})}\over{dM}} 
b(M,z_{in}) y_l(M,z)\right]}
^2
\ee
where $b(M,z)$ is the time dependent linear bias factor. The matter power 
spectrum, $P_m(k,z_{in})$, is related to the galaxy correlation function 
$P_g(k,M1,M2,z_{in})$ through the bias i.e
$P_g(k,M1,M2,z_{in})=b(M1,z_{in})b(M2,z_{in})D^2(z_{in})P_m(k,z=0)$ where we 
adopt $b(M,z)$ given by
$b(M,z) = (1+0.5/\nu^4)^{0.06-0.02n}(1 + (\nu^2 -1)/\delta_c)$ (Jing 1999).
This expression
for the bias factor matches accurately the results of
 N-body simulation for small halos .
This fitting formula for bias does not underestimate the
clustering of small halos with $\nu < 1$, and accurately fits simulation
results of CDM models and scale-free models. The difference between the
fitting formula and the simulation result is generally less than $\sim
15\%$.
In the above equation $D(z)$ is the linear growth factor of density
fluctuation, $\delta_c=1.68$ and $\nu=\delta_c / \sigma(M)$. In Eqn 6., we have
utilized Limber approximation (Limber 1954) and have set $k=l/r_z$, where
$r_z$ is the comoving angular diameter distance.

\bigskip
\section{Results and discussions}

We plot the mean Compton 
distortion due to thermal and temperature distortion due to kinetic SZ effects 
as a function of $z_{in}$ in Figure 2. As can be seen from the figure,
wind originating at redshifts $\sim 6\hbox{--}8$
distort the CMB more than
those which may have originated relatively earlier, or those more recent. This 
can be naively understood as follows: 
the distortion of the CMB due to winds is proportional to the number of galactic
winds originating at $z_{in}$;
 the distribution of the galaxies follow
from the Press-Schecter formalism where the number density per comoving volume 
peaks at
redshift $\sim 5$ for the mass range relevant here, 
and the falls off at higher redshifts. 
This combination produces the maximum
distortion at an intermediate value of $z_{in}$. 
%Comparing the distortions
%with the COBE limit of $y \le 1.5 \times 10^{-5}$ (horizontal 
%line in the figure), 
%we see that models of galactic winds originating at $4.5 \le z_{in} \le 9$ are
%ruled out. Changing $v_r(z=0)$ to $300$ km s$^{-1}$ makes the violation
%of the COBE limit only marginal, at $z_{in} \sim 6\hbox{--}7$.

%We note that this range of redshift has been considered to be a possible
%epoch of galactic winds enriching the IGM in the recent simulations
%by Aguirre \etal (2000), although they did not discuss the constraints
%from CMB.

We also show in Figure 2 our results for an the case of $M_b=0.05 \, M$ 
(thin lines), which
represents the case of winds with efficiencies decreased by a factor of two.
%It is seen that the wind models marginally violate the COBE limit for
%$z_{in} \sim 6\hbox{--}7$ (for $v_r(z=0)$ to $400$ km s$^{-1}$).

Next, we focus on the angular power spectrum of temperature anisotropy from
galactic outflows. First, we plot in Figure 3 the anisotropy due to kinetic SZ
effect from outflows
with and without clustering. We plot the power
spectra for outflows for three initial redshift $z_{in}=7, 9, 11$.
%Note that the case for $z_{in}=7$  violates the $y$-parameter
%constraint 
%(and is given only for comparison) while the other cases of $z_{in}$ 
%plotted in this figure do not.
For $z_{in}=9$, we present the results for 
$v_r(z=0)=400$ and $300$ km s$^{-1}$.
We also plot in Figure 3 the angular power spectra due to
thermal SZ effect from clusters of galaxies, with and without clustering,
for comparison.

It is evident from Figure 3, that the power spectrum due to clustering of the 
sources (thick lines) can
be very important from CMB distortions due to winds at high redshifts. 
In comparison,
for clusters of galaxies, the Poisson power 
spectrum is larger than the clustering power spectrum, 
as shown by Komatsu \& Kitayama
(1999). In that case,
if flux limited clusters detected in X-ray 
surveys are subtracted, then 
clustering power spectrum can dominate at $l<200$. 
In all the cases that we consider here, however, we 
find that clustering power spectrum is dominant below $l \sim 3 \times 10^4$.
This is easy to understand in the following manner.
Since, $C_l^{clustering} / C_l^{Poisson} \sim n(M,z)b^2(M,z) D^2(z) P_m (k) $,
with $k \sim l/r(z)$,
the bias increases at high $z$ along with $r(z)$, thereby
boosting up the clustering power 
spectrum. The evolution of bias with redshift has been 
studied by many authors both
theoretically (Tegmark \& Peebles 1998, Valageas, Silk \& Schaeffer 2000)
and through simulations (Blanton \etal 1999, Dav\'e \etal 1999). All of these
studies show that the bias increases rapidly with increasing redshift.

The thick lines in Figure 3 also show that the clustering spectra
from winds are greater than the
clustering spectrum from clusters of galaxies above $l \sim 10^3$ in almost
all cases we consider. The clustering spectra from winds are , however,
in general smaller than the Poissonian
spectrum from clusters of galaxies.

The thin lines in 
Figure 3 show that the Poissonian spectra from winds (kinetic SZ)
peak at $l \sim 4 - 6 \times 10^4$.
Although it is swamped by the thermal SZ signature from galaxy clusters,
we note that the frequency dependence of kinetic SZ effect is different
from that of the thermal SZ effect. So, it would be possible in principle
to separate out this important signature from winds.
This angular scale is an order of magnitude less than the peak for clusters of
galaxies. This is beyond the ranges of observations with the upcoming PLANCK
satellite, but is well within the range for ALMA. However, it may well be
swamped by other secondary distortions of the CMB at sub arc-min scales.

Recently, observational upper limits (shown in figure 3 with open circle)
have been put on the arc-minute scale distortion
of the CMB temperature anisotropy by ATCA (Subrahmanyan \etal 1998) and 
BIMA (Holzapfel 
\etal 1999). In the R-J limit $\Delta T_l = T_{CMB}\sqrt{l(l+1)C_l}$.
Angular power spectrum from any feasible model must satisfy these upper limits
in addition to the limit on the mean y-distortion. 
This has been used for the thermal SZ 
effect from hot gas in galaxy clusters to constrain
cosmological models (Majumdar \& Subrahmanyan 2000). 
None of our wind models, however, violate the ATCA and BIMA limits,
and hence cannot be
ruled out independent of the violation of COBE limit on $y$.

The result of the power spectrum due to clustering dominating the anisotropy
brings about an interesting possibility to probe bias at high redshift.
Komatsu \& Kitayama (1999) have noted that a possible detection of the 
clustering spectrum due to galaxy clusters would give us information
on bias at high redshift. In our case, we find that the clustering spectrum
due to winds (kSZ) is much larger than the clustering spectrum
from galaxy clusters (tSZ), and smaller than the Poissonian spectrum from
galaxy clusters (tSZ), for a wide range of $l$. If the X-ray luminous
clusters are removed, then it may be possible to detect the
spectrum due to clustering of outflowing galaxies. Moreover, the different
frequency dependence of kinetic and thermal SZ effects may allow us
to separate these two effects. This may give valuable information about
the evolution of bias at high redshifts, although we note that the
parameters for the wind shells are not very certain, as we explain below.

We then plot the power spectra due to thermal SZ effect from winds
in Figure 4. As the figure suggests, the thermal SZ effect is much smaller than
the kinetic SZ effect in both Poissonian and clustering cases. It is also
much smaller than that from clusters of galaxies.

We compare the results with different efficiency of winds in Figure 5,
where we show the results for the cases $M_b=0.1 \, M$ (thick lines) 
and $M_b=0.05 \, M$ (thin lines). The curves show how a change of a
factor of two in the assumption of wind efficiencies would change our results.

In figure 6, we have plotted the power spectrum for both tSZ and kSZ
distortions from winds for three representative frequencies. We have
chosen
to show the results for 100, 217 and 545 GHz. These are 3 of the proposed
9 frequencies in which the Planck surveyor satellite mission would
operate. The larger 2 frequencies would have a resolution of about $5'$, and
so would be able to go to $l>2000$. Note that at these $l$ values the
primary anisotropy drops and the power spectrum would be dominated by
secondary anisotropies. Also, the clustering power spectra due to
distortion by winds peak at these angular scales, as evident in figure 3.
At 100 GHz and 545 GHz, the effect from tSZ distortion due to
clusters of galaxies is greater than that from winds. However, we can take
advantage of the crossover in the frequency dependence of tSZ, as given in
$g(x)$, where the contribution from tSZ goes to zero. The middle panel in
figure 6, shows the spectra at 217 GHz. At this frequency the tSZ from
both cluster of galaxies and winds is zero, and we see that the
clustering spectra due kSZ from winds exceeds the Poisson spectra from kSZ
from galaxy clusters. This difference is further
highlighted in figure 7, where we plot the composite spectra (i.e sum of
both Possion and clustering spectra) for distortion due to galaxy clusters
and wind at 217 GHz. It is clear that at this frequency, distortion due to
wind peaks at $l \sim 2000-3000$ and is an order of magnitude  greater
than distortion due to galaxy clusters. We can thus hope to have an
attempt at detecting SZ distortion due to galactic winds at high redshift
with Planck.
We also note that the anisotropies are likely to be detected and measured
in the future by the proposed long base line interferometers such as ALMA.

The above results, however, should be viewed in light of the uncertainties
inherent in our calculation. Firstly, the parameters of the outflows are by
no means certain. The mechanical luminosity of galaxies, for example, 
is somewhat uncertain. The efficiency of galactic outflows, and its dependence
on the mass and redshift, are also not well known. At any rate, the curves
in Figure 5 gives an idea of the magnitude of changes that might occur
if some of these assumptions are changed. Lastly, our assumption
for a single epoch of galactic outflows, might be naive. Also, we have 
assumed a mean peculiar velocity for all galaxies. Perhaps a simulation
with a distribution in $z_{in}$ and $v_r$ would be able to address these
issues in a better way. 

We also note here that we have neglected the anisotropy caused by the
inverse Compton cooling of the shells of outflows.  Also, we have not
taken into account the heating of the gas interior due to collision of
the shell with the IGM. Our estimates of the
anisotropy is, therefore, a conservative one.

Our results, however, seem to be robust with regards to the assumed upper mass
limit for the outflowing galaxies, as long as the the upper limit is
greater than $\sim 10^9$ M$_{\odot}$. We note that Scannapieco
\& Broadhurst (2000) have calculated the upper mass limit of outflowing
galaxies in the $\Lambda$CDM universe at $z \sim 10$ to be of order
$10^9$ M$_{\odot}$. Our results do not change significantly if we assume
this upper mass cutoff instead of the one described earlier.

As far as cosmologies other than the $\Lambda$CDM are concerned,
we found that the resulting anisotropy is less in the case of a
sCDM ($\Omega =1, \Lambda=0, h=0.5$) 
universe for a given $z_{in}$. This is mainly because of the fact that 
structures form later in sCDM universe although the value of $\sigma_8$
is larger in this case, and since the distortion in the
case of wind is biased towards the high redshift. The anisotropy in 
the case of the OCDM universe ($\Omega=0.35, \Lambda=0 ,h=0.65$)
 is slightly less than in the $\Lambda$CDM
universe. There are two reasons for this: in the OCDM universe, firstly
the COBE normalised $\sigma_8$ for OCDM is much less than in $\Lambda$CDM 
case, and secondly, peculiar velocities are also smaller
at high redshift, for a given normalization at the present epoch.

%It is, however, interesting to note that CMBR observations can be used
%to constrain the epoch of widespread galactic outflows at high redshift.
%It is also interesting that the allowable range of $z_{in}$ for winds
%($z_{in} \ge 9$ for $\Lambda$CDM universe)
%has been shown to be the range of occurrence of first luminous objects in the
%universe, close to the epoch of reionization (e.g., Gnedin \& Ostriker 1997).

\bigskip
\section{Conclusions}

We have calculated the SZ distortion of the CMB due to galactic winds
at high redshift, originating at a single epoch $z_{in}$, from galaxies
of masses between $5 \times 10^7$ and $10^{11}$ M$_{\odot}$, in a 
$\Lambda$CDM universe. 
We summarise our findings below:

(a) We confirm the previous
estimates of the mean $y$-distortion due to thermal SZ effect. We, however,
found the kinetic SZ effect to be more important than the thermal SZ effect 
in terms of the effect on the angular power spectrum of distortions.
%, and that
%the COBE limit on the $y$-parameter is violated by galactic winds originating
%at $4.5 \le z_{in} \lsim 9$. This result does not change significantly
%if the upper mass cutoff is decreased to $\sim 10^9$ M$_{\odot}$. If
%the normalization of peculiar velocities at present epoch is decreased to
%$v_r (z=0)=300$ km s$^{-1}$ then the COBE limit is margly violated at
%$z_{in} \sim 6\hbox{--}7$. 
%The COBE limit is not violated in the case of a OCDM or sCDM universe.
%We have noted that 
%this range of redshift has been considered in recent simulations of
%enrichment of the IGM by galactic winds. we have also noted that if the
%wind efficiencies are decreased by a factor of two, then the COBE limit
%is marginally violated for $z_{in} \sim 6\hbox{--}7$.

(b) We obtained the angular power spectrum  of
distortions with and without clustering of parent galaxies. The power spectrum
due to kinetic SZ effect (Poisson) is found to be comparable or larger than
the SZ effect from clusters of galaxies for $l \gsim  10^5$.

(c) We found that clustering of low mass galaxies at high redshift
could increase the angular power spectum of distortions. 
The power spectrum due to
clustering of parent galaxies of outflows (kinetic SZ) was found to be 
also larger
than the clustering power spectrum from galaxy clusters (thermal SZ)
and somewhat smaller than the Poisson power spectrum from galaxy clusters
(thermal SZ). We have explicitly shown the frequency dependence of the
various power spectra. 
It is possible that the clustering power spectrum from winds
can be estimated after subtracting X-ray luminous clusters dominating
the Poisson power spectrum from clusters, or by using the different frequency
dependence for thermal and kinetic SZ effects, yielding information on 
bias of low mass galaxies at high redshift. 

(d) We have shown how the total power spectra (kinetic and thermal SZ,
including both Poisson and clustering effects) for winds will dominate
over the corresponding spectra for clusters, at $217$ GHz, a proposed
frequency for the PLANCK satellite mission.

We conclude that the SZ distortion from galactic winds at high
redshift, if present,
could constitute an important source of secondary CMB anisotropy on arc
minute and sub-arcminute scales.

\bigskip
\section{Acknowledgement}

The authors would like to thank the referee, Priyamvada Natarajan, for her 
detail comments which
have improved the paper considerably.
SM would like to thank the Raman Research Institute, Bangalore, for
hospitality and for providing computational facilities. 
BN and MC acknowledge a grant from the India-Japan Cooperative Research
Programme for partial support of this work and BN acknowledges the
hospitality of the National Astronomical Observatory, Tokyo.
SM also wishes to thank Pijush Bhattacharjee for all the encouragements.

%%%%%%%%%%%%%%%%%%%%%%%%%%%%%%%%%%%%%%%%%%%%%%%%%%%%%%%%%%%%%%%%%%%%%%%%%%%%

%%%%%%%%%%%%%%%%%%%%%%%%%%%%%%%%%%figures%%%%%%%%%%%%%%%%%%%%%%%%%%%%%%%%%%%

\begin{figure}
\figurenum{1}
\plotfiddle{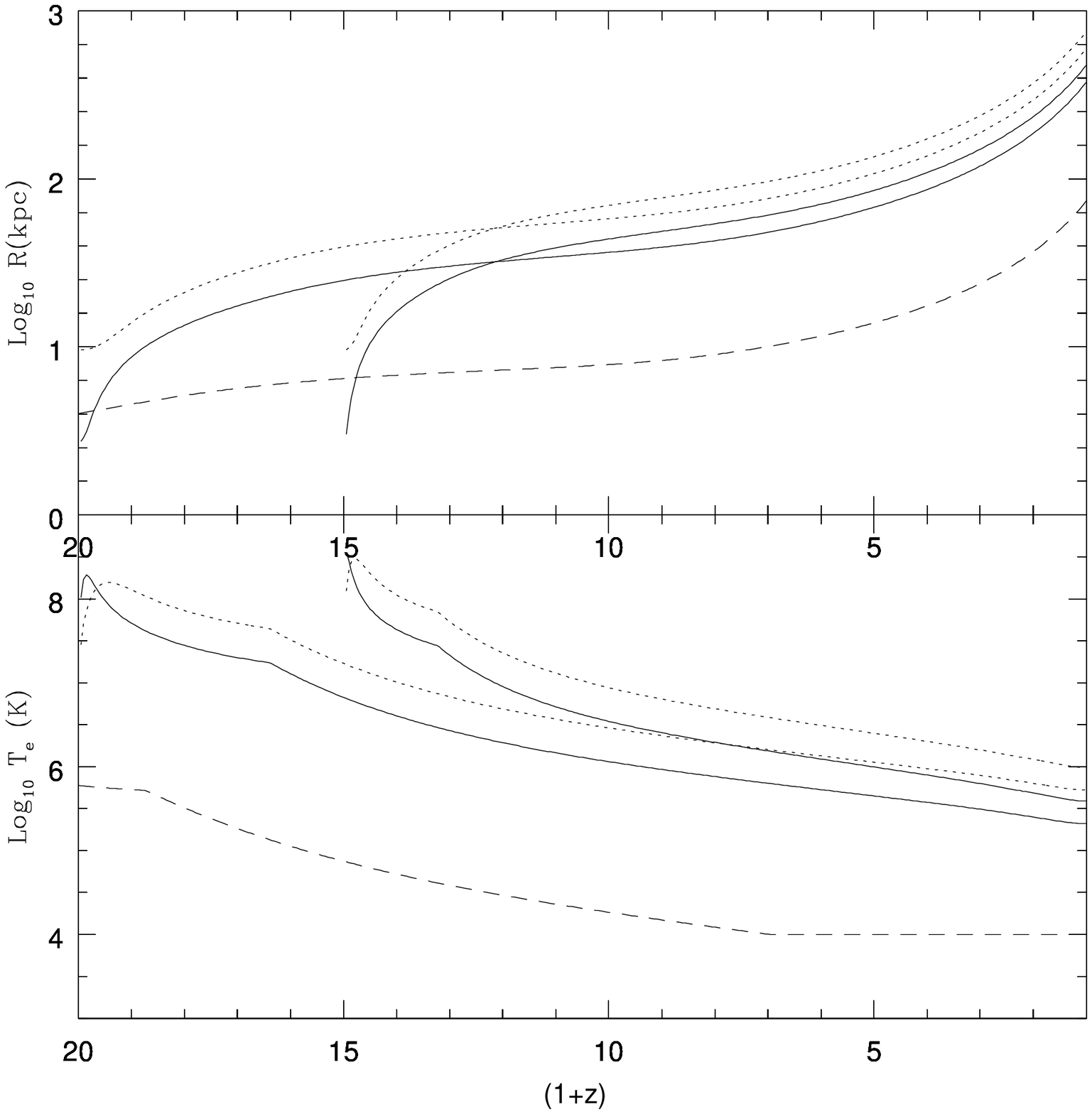}{5.0in} {0.0}{70.0}{70.0}{-225.0}{-50.0}
\caption {The evolution of radius and electron temperature are shown for
$1+z_{in}=15$ and $20$, and for two masses of galaxies, $M=10^9$ (solid
lines) and $10^{10}$ M$_{\odot}$ (dotted lines).
We also plot with dashed lines the case for wind for $M=5 \times 10^7$
M$_{\odot}$, with $z_{in}=25$, with the parameters used by Scannapieco
\& Broadhurst (2000), for comparison. The temperature is not allowed
to fall below $10^4$ K, as in their work, considering the heating
due to UV background radiation.
}
\end{figure}

\clearpage

\begin{figure}
\figurenum{2}
\plotfiddle{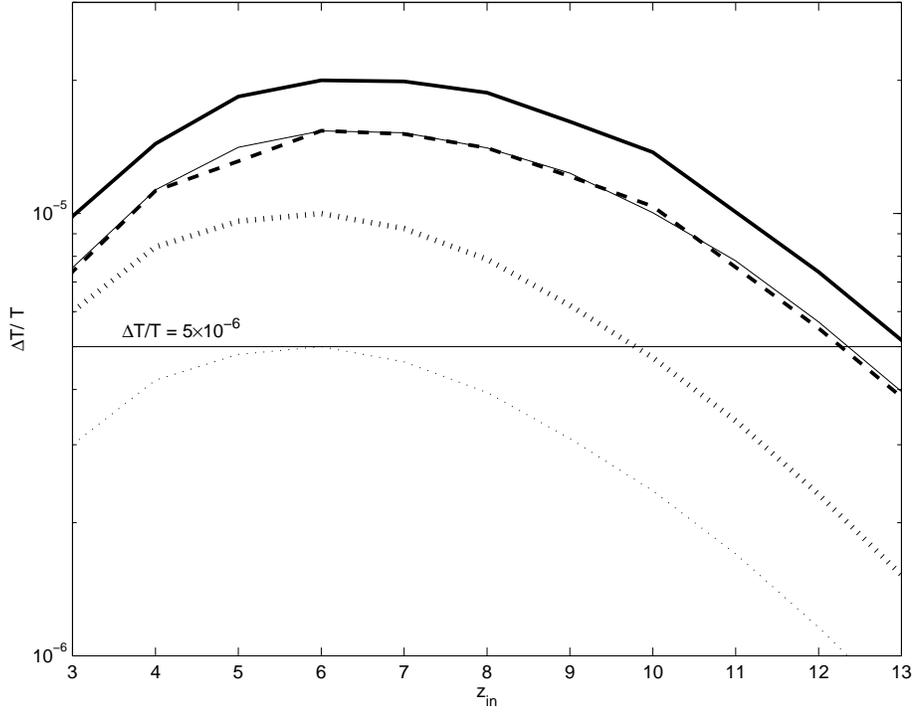}{5.0in} {0.0}{70.0}{70.0}{-225.0}{-50.0}
\caption {The temperature distortion ${\Delta T}\over T$ (in the R-J limit) is
plotted against $z_{in}$ for both kSZ and tSZ effect due to winds. The
thick solid curve is for kSZ with $v_r(z=0)=400 km s^{-1}$ and the thick
dashed curve is for $v_r(z=0)=300 km s^{-1}$. Both are with $M_b=0.1M$.
The thin solid curve is for kSZ with $v_r(z=0)=400 km s^{-1}$ and
$M_b=0.05M$. The dotted curves are for tSZ with the solid one for
$M_b=0.1M$ and the thin one for $M_b=0.05M$. A horizontal line with
${\Delta T \over T} = 5\times10^{-6}$ is shown for easy referencing.
}
\end{figure}

\clearpage
\begin{figure}
\figurenum{3}
\plotfiddle{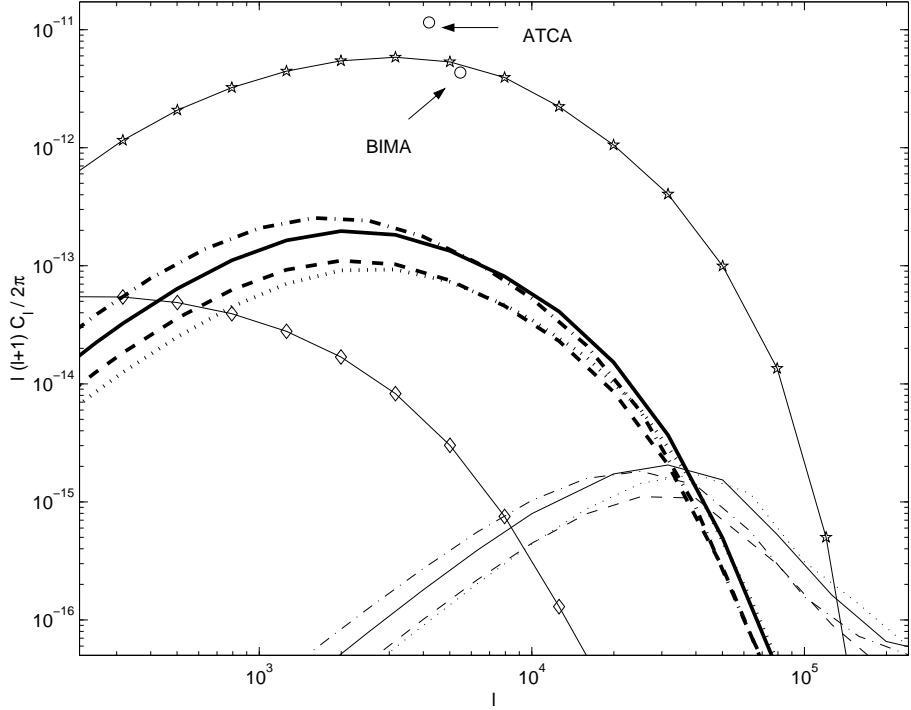}{5.0in} {0.0}{70.0}{70.0}{-225.0}{-50.0}
\caption {Angular Power spectrum due to kinetic SZ effect from winds are shown for
different models. The thick lines show clustering power spectrum and the thin
lines show Poisson power spectrum. The models shown are : (1) $z_{in}=9,
v_r(z=0)=400$ km s$^{-1}$ (solid line), (2) $z_{in}=9,v_r(z=0)=300$ km s$^{-1}$
(dashed line), 3) $z_{in}=7,v_r(z=0)=400$ km s$^{-1}$ (dot-dashed line) and
$z_{in}=11,v_r(z=0)=400$ km s$^{-1}$ (dotted line). 
For comparison, Poisson (solid
line with stars) and clustering (solid line with squares) spectra from 
from thermal SZ effect from galaxy clusters are shown. We have used the same
cosmological model as in Komatsu \& Kitayama (1999)  
for comparison (with their figures 1 \& 4).
The ATCA abd BIMA upper
limits are also marked in the figure by the two open circles.
}

\end{figure}

\begin{figure}
\figurenum{4}
\plotfiddle{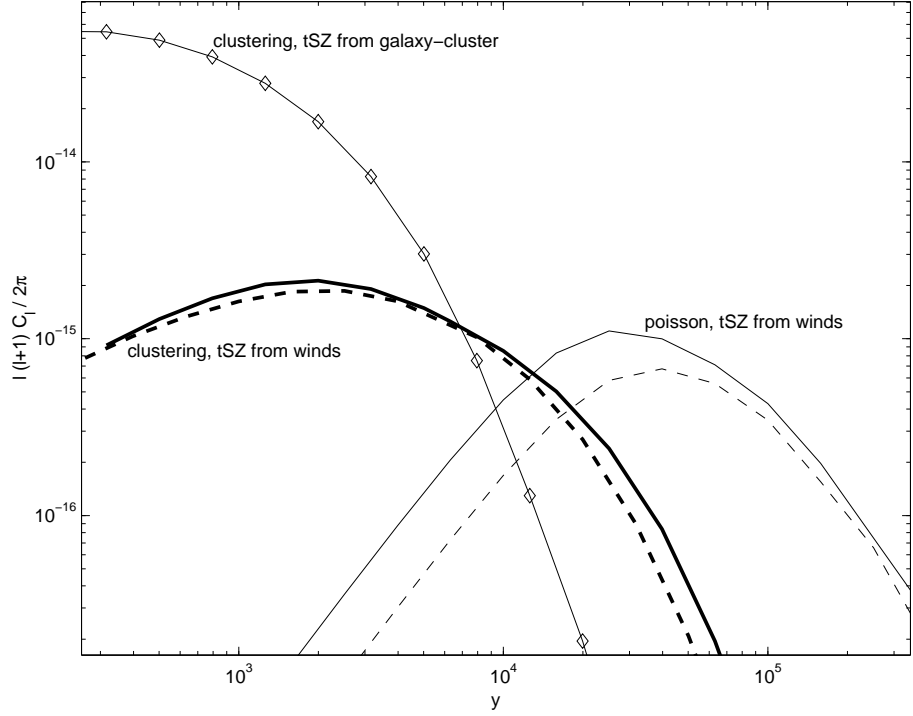}{5.0in} {0.0}{70.0}{70.0}{-225.0}{-50.0}
\caption {Angular Power spectrum due to thermal SZ effect from winds 
are shown for different models.
The thick lines show clustering power spectrum and the thin
lines show Poisson power spectrum. The models shown are : (1)$z_{in}=9$,
(solid line), (2) and
$z_{in}=11$ (dashed line).For comparison clustering 
(solid line with squares) spectrum from 
from thermal SZ effect from galaxy clusters are shown.
}
\end{figure}

\begin{figure}   
\figurenum{5}
\plotfiddle{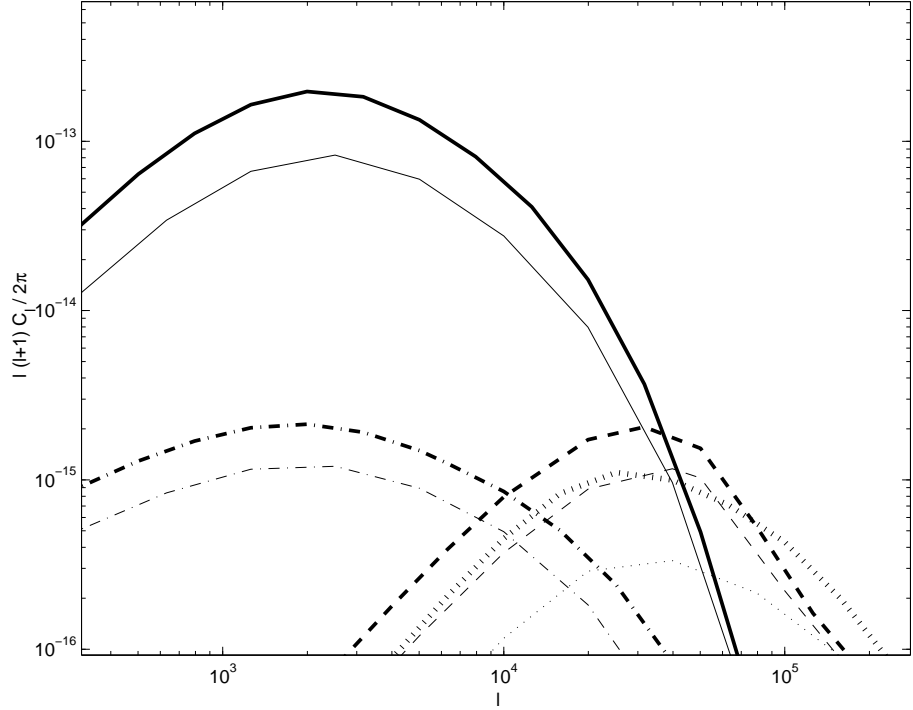}{5.0in} {0.0}{70.0}{70.0}{-225.0}{-50.0}
\caption {
 A comparison of the power spectrum for the two
cases (i) $M_b=0.1M$ ( thick lines) and (ii)$M_b =0.05M$ (thin lines) is
shown, for $z_{in}=9$ and $v_{ro}=400 km s^{-1}$. 
The solid lines refer to clustering spectra due to kSZ; the dashes
lines refer to the Poisson spectra for kSZ; the dot-dashed lines refer to
the clustering spectra for tSZ ;and the dotted lines refer to the Poisson
spectra for tSZ.
}
\end{figure}

\begin{figure}   
\figurenum{6} 
\plotfiddle{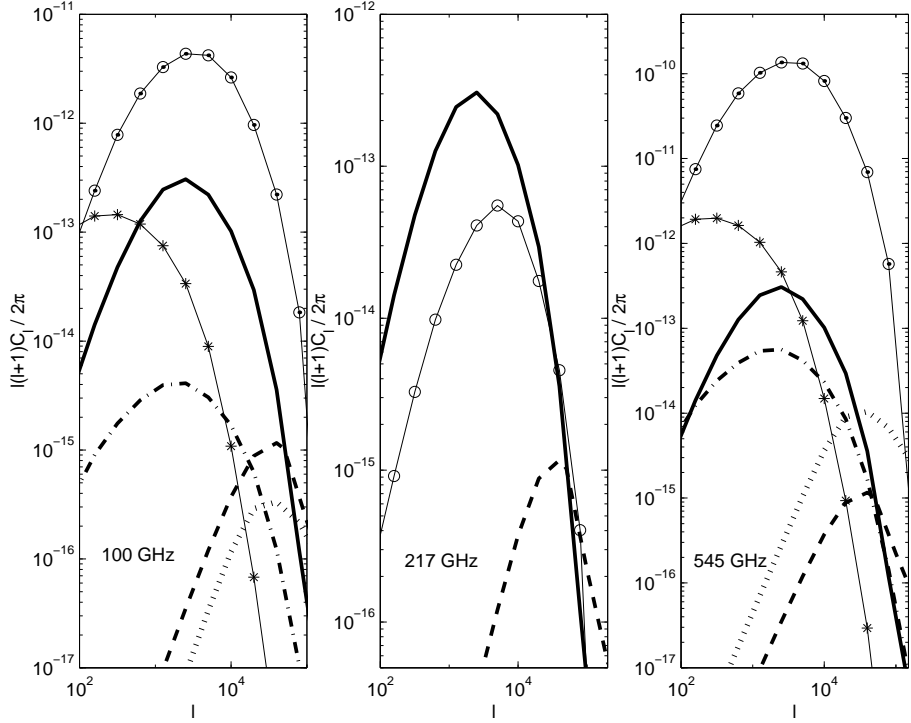}{5.0in} {0.0}{70.0}{70.0}{-225.0}{-50.0}
\caption {
The figure shows the power spectrum at different frequencies. The
thick lines in all the subplots are due to winds with $z_{in}=9$ ;  solid
linea are for clustering spectra due kSZ with $v_{r,0}=400 km s^{-1}$ and
the dashed lines are the Poisson spectra for the same; the dot-dashed
lines are for Clustering spectra due to tSZ from winds and the dotted
lines are the Poisson spectra for the same. For comparison, in the left and
the right figures, Poisson spectra for tSZ from clusters (lines with filled
circles) and clustering spectra for the same (lines with stars) are shown.
In the middle panel, the line with open circle shows the Poisson spectra
due to kSZ from galaxy clusters.
} 
\end{figure}

\begin{figure}
\figurenum{7}
\plotfiddle{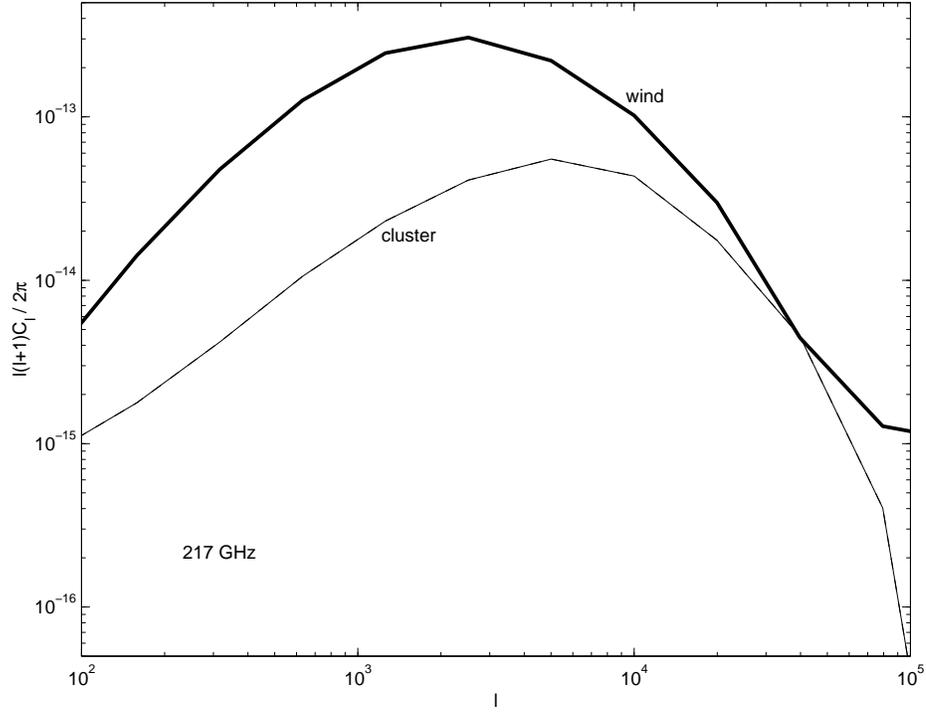}{5.0in} {0.0}{70.0}{70.0}{-225.0}{-50.0}
\caption {
The composite power spectra (kSZ+tSZ, including contribution from both
clustering and Poisson spectra) is shown for distortion from winds (thick
line) and clusters (thin line), $z_{in}=9$ and $v_{ro}=400 km s^{-1}$. 
The spectra are calculated for the
proposed Planck observation frequency of 217 GHz.
}
\end{figure}

\end{document}